\documentclass[a4paper,fleqn]{cas-sc}

\usepackage[numbers]{natbib}

\usepackage{amsmath,amssymb}
\usepackage{graphicx}
\usepackage{booktabs}
\usepackage{subcaption}

\graphicspath{{./Figures/}}

\begin{document}
\let\WriteBookmarks\relax
\def\floatpagepagefraction{1}
\def\textpagefraction{.001}

\shorttitle{Analog realization of a forced Aliev--Panfilov oscillator}
\shortauthors{Z. Fu and U. Pillai}

\title[mode=title]{Analog Realization and Measurement of a Forced Aliev--Panfilov
Excitable Oscillator}

\author[1]{Zixuan Fu}
\cormark[1]
\ead{zf2375@nyu.edu}
\credit{Methodology, Investigation, Formal analysis, Data curation,
Visualization, Writing -- original draft}

\affiliation[1]{organization={Department of Electrical and Computer Engineering,
            New York University Tandon School of Engineering},
            city={Brooklyn},
            state={NY},
            postcode={11201},
            country={USA}}

\author[1]{Unnikrishna Pillai}
\ead{up240@nyu.edu}
\credit{Conceptualization, Supervision, Writing -- review \& editing}

\cortext[1]{Corresponding author}

\begin{abstract}
A single resistor moves a discrete analog realization of the forced Aliev–Panfilov excitable oscillator from a chaotic operating point of the model to a regular one. In this model a rate parameter multiplies the complete right-hand side of the excitation equation, forcing term included. The input resistor of the excitation integrator sets that parameter, so no state, delay element, or controller is added. Sampled waveforms were recorded at two resistor values, with two independent runs each. The measured excitation excursion agrees with numerical integration to within three percent at the baseline setting and five percent at the rate-controlled setting. Between the two settings the excitation excursion contracts by a factor of thirteen. At the rate-controlled setting the recovery excursion exceeds prediction by a factor of eight. This excess traces to a displaced operating point on the state-dependent recovery rate. Thermal drift in the multiplier chain and leakage at the high-impedance recovery-rate divider could each produce it, and the present data do not separate them. A tested bipolar input stage strongly loads that divider, a constraint absent from the ideal equations. The divider, not the sub-hertz signal bandwidth, sets the amplifier input-stage requirement.
\end{abstract}

\begin{highlights}
\item One integrator resistor moves an analog excitable oscillator from chaos to regularity
\item The resistor scales the whole excitation right-hand side, forcing term included
\item Excitation excursion matches numerical integration within 3\% and 5\% at two settings
\item The recovery-rate divider, not signal bandwidth, sets the op-amp input requirement
\item A displaced recovery operating level explains the excess recovery excursion
\end{highlights}


\maketitle

\section{Introduction}\label{sec:intro}

Forced excitable oscillators serve as models of cardiac and neural dynamics. They
also serve as building blocks of nonlinear and neuromorphic circuits. Compact
chaotic circuits have long been attractive because a few analog components can
generate rich dynamics~\cite{piper,sprott,boriskov}. Practical oscillators often
need an equally simple way to leave an unwanted chaotic regime. Classical
chaos-control methods usually add a control mechanism for this purpose. OGY
control applies parameter perturbations synchronized to a Poincar\'e
section~\cite{ogy}. Pyragas control requires a delay matched to a target unstable
periodic orbit~\cite{pyragas}. Dedicated feedback has also been used to
stabilize nonlinear oscillator circuits~\cite{tsubone}. These methods suit cases
where an embedded orbit must be preserved. They require a tuned loop, a delay
element, or extra control logic. Recent work continues to add designed
controllers. One example is an adaptive law that drives an uncertain Lorenz
system to rest in an arbitrary prescribed time~\cite{wang2024}.

This paper shows that a single passive resistor is enough to take a forced
analog excitable oscillator out of its chaotic regime. The oscillator is the
forced Aliev--Panfilov model, realized as a real-time analog computer on one
breadboard panel. A rate parameter of the model multiplies the \emph{entire}
right-hand side of the excitation equation, with the forcing term included. In
the circuit, the input resistor of the excitation integrator alone sets this
parameter. Replacing that resistor of $1$\,M$\Omega$ with one of $6$\,M$\Omega$
moves the board from an operating point where the model is chaotic to one where
the model is regular. The largest Lyapunov exponent of the model changes sign
from positive to negative between the two settings. On the bench, the measured
excitation excursion contracts by a factor of $12.9$, against $13.9$ predicted,
and the orbit locks to the forcing period. No state, delay element, sensed
feedback, or digital controller is added. In the model, the regular setting also
tolerates a $\pm5\%$ error in the resistor and $\pm15\%$ errors in the forcing
amplitude and frequency.

The contributions of the paper are the following.
\begin{itemize}
\item A hardware realization of chaos control by one resistor. The input
resistor of the excitation integrator switches the forced Aliev--Panfilov
oscillator between the chaotic and the regular operating points of the model. No
controller dynamics are added.
\item A measured validation of the realization. Sampled records from two
independent runs at each setting are compared with numerical integration of the
governing equations. The sampled records are provided with the paper.
\item An amplifier design rule absent from the ideal equations. The
high-impedance recovery-rate divider, not the sub-hertz signal bandwidth, sets
the input-stage requirement of the amplifiers.
\item A localization of the one remaining discrepancy. The excess recovery
excursion at the rate-controlled setting is traced to a displaced operating level
on the state-dependent recovery rate.
\end{itemize}

Section~\ref{sec:circuit} derives the circuit from the governing equations,
fixes the component values, and identifies the amplifier input-stage
requirement. Section~\ref{sec:meas} describes the prototype and the measurement
protocol, and it reports the measured behavior at both settings.
Section~\ref{sec:analysis} analyzes the deviation in the recovery variable, the
margin of the rate-controlled setting, and the scope of the experiment.
Section~\ref{sec:concl} summarizes the contributions.

\section{Circuit Realization}\label{sec:circuit}

\subsection{Governing Equations and Operating Points}\label{sec:model}

The circuit in Fig.~\ref{fig:schematic} realizes the forced Aliev--Panfilov
model~\cite{aliev}. The state variables are the excitation variable $u$ and the
recovery variable $v$. Both are dimensionless, and time $t$ is in seconds. The
model reads
\begin{align}
\frac{du}{dt} &= \frac{1}{1+\gamma}\Big(-K u(u-a)(u-1)-uv+F\cos\omega t\Big),
  \label{eq:u}\\
\frac{dv}{dt} &= \varepsilon(u,v)\Big(-v-K u\,(u-a-1)\Big),
  \label{eq:v}
\end{align}
with the state-dependent recovery rate
\begin{equation}
\varepsilon(u,v)=\varepsilon_0+\frac{\mu_1 v}{u+\mu_2}.
\label{eq:eps}
\end{equation}
The constants $K$, $a$, $\varepsilon_0$, $\mu_1$, and $\mu_2$ set the excitation
and recovery kinetics. The forcing has amplitude $F$ and angular frequency
$\omega$. Table~\ref{tab:params} lists their values. The rate parameter
$\gamma\ge0$ multiplies the entire right-hand side of Eq.~\eqref{eq:u}, forcing
term included. It adds no term to the vector field. The baseline setting is
$\gamma=0$. The rate-controlled setting uses a nominal $\gamma=5$. Measured
resistor values give a realized $\gamma=4.881$, written $4.88$ in the figures
and tables.

The largest Lyapunov exponent $\lambda_1$ of Eqs.~\eqref{eq:u}--\eqref{eq:eps} is
\begin{equation}
\lambda_1=\lim_{T\to\infty}\frac{1}{T}\ln\frac{\lVert\boldsymbol{\delta}(T)\rVert}
{\lVert\boldsymbol{\delta}(0)\rVert},
\label{eq:lyap}
\end{equation}
where $\boldsymbol{\delta}(t)$ is a perturbation of the state $(u,v)$ that
evolves under the linearized equations. A long-horizon Benettin calculation with
fourth-order Runge--Kutta integration gives $\lambda_1$. Table~\ref{tab:params}
lists the step, transient, observation window, and renormalization interval. At
$\gamma=0$ the value of $\lambda_1$ depends on the initial condition. It is
positive for all six initial conditions tested, as Table~\ref{tab:params}
shows. At the realized $\gamma=4.881$, $\lambda_1$ is negative. The model is
therefore numerically chaotic at the baseline setting. At the rate-controlled
setting it is attracting in the physical state variables.

\begin{table}[pos=!htbp]
\caption{Model parameters, circuit values, computed largest Lyapunov exponents,
and acquisition settings.\label{tab:params}}
\centering
\begin{tabular}{ll}
\toprule
Quantity & Value \\
\midrule
$K$, $a$ & $8$, $0.12$ \\
$\varepsilon_0$, $\mu_1$, $\mu_2$ & $0.01$, $0.2$, $0.3$ \\
Forcing $F$, $\omega$ & $0.40$, $1.8$\,rad\,s$^{-1}$ \\
\midrule
State scaling & $u=V_U/5$, $v=V_V/2$ \\
$C_u$, $R_0$, $R_f$ & $1\,\mu$F, $1$\,M$\Omega$, $100$\,k$\Omega$ \\
Multiplier scale factor & $1/10$ \\
\midrule
Integration step, transient & $0.01$\,s, $200$\,s \\
Observation window, renormalization interval & $1500$\,s, $0.5$\,s \\
$\lambda_1$ at $\gamma=0$, range over six initial conditions & $+0.113$ to $+0.132$\,s$^{-1}$ \\
$\lambda_1$ at $\gamma=0$, mean over six initial conditions & $+0.125$\,s$^{-1}$ \\
Range of initial conditions & $u\in[0,0.8]$, $v\in[0,1.5]$ \\
$\lambda_1$ at $\gamma=4.881$ & $-0.0419$\,s$^{-1}$ \\
\midrule
Generator frequency, amplitude, offset & $286.500$\,mHz, $4.000$\,V$_{\mathrm{pp}}$, $0$\,V \\
Oscilloscope mode & roll, $20$\,s/div \\
Record length & $6400$ points over $200.0$\,s \\
Forcing periods per record & $57.3$ \\
\bottomrule
\end{tabular}
\end{table}

\subsection{Analog-Computer Topology}\label{sec:topology}

\begin{figure}[pos=htbp]
\centering
\includegraphics[width=\linewidth]{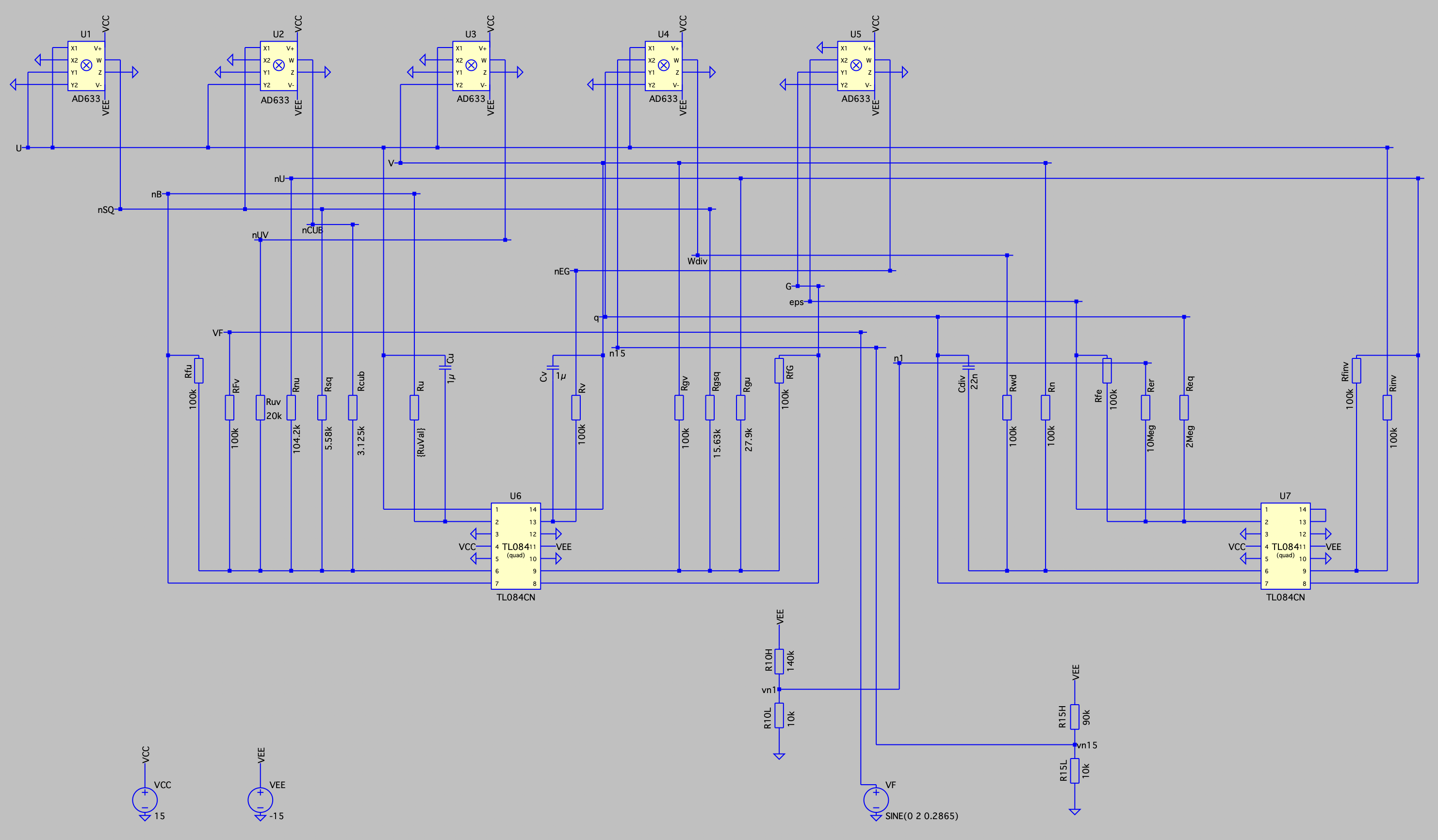}
\caption{Analog-computer realization of Eqs.~\eqref{eq:u}--\eqref{eq:eps}. Two
TL084 JFET-input quad operational amplifiers implement the integrators, summers,
and division stage. Five AD633 four-quadrant multipliers implement $u^2$, $u^3$,
$uv$, the divider loop, and the recovery-rate product. The integrator input
resistor $R_u$ alone sets the rate parameter $\gamma$. A value of
$1$\,M$\Omega$ gives $\gamma=0$, and $6$\,M$\Omega$ gives $\gamma=5$.}
\label{fig:schematic}
\end{figure}

The circuit follows from Eqs.~\eqref{eq:u}--\eqref{eq:eps}. The equations are
first recast as a block diagram of integrators, summers, and multipliers.
Component values then set the parameters. The node voltages $V_U$ and $V_V$ at
nodes U and V of Fig.~\ref{fig:schematic} carry the states, with
\begin{equation}
u=\frac{V_U}{5},\qquad v=\frac{V_V}{2}.
\label{eq:scale}
\end{equation}
The state ranges $u\in[-0.2,1]$ and $v\in[0,2.5]$ then fit within the $\pm15$\,V
supply rails with margin. The $u$ integrator has input resistor $R_u$ and
capacitor $C_u$, with
\begin{equation}
R_u=R_0(1+\gamma).
\label{eq:Ru}
\end{equation}
With $R_0C_u=1$\,s, the integrator gain $1/(R_uC_u)$ equals $1/(1+\gamma)$ per
second. The circuit therefore runs in real time and realizes Eq.~\eqref{eq:u}
directly. The forcing frequency $\omega=1.8$\,rad\,s$^{-1}$ corresponds to
$0.2865$\,Hz at the signal generator.

Fig.~\ref{fig:schematic} shows the analog computer, built around five AD633
four-quadrant multipliers. Three of them form $u^2$, $u^3$, and $uv$. One takes
part in the division loop, and one forms the recovery-rate product. A summing
amplifier assembles the nonlinear terms \emph{together with} the forcing input.
The summed node drives the $u$ integrator through $R_u$ alone. Because the
forcing joins the sum upstream of $R_u$, a change in $R_u$ scales the complete
right-hand side of Eq.~\eqref{eq:u}, not the nonlinearity only. The
single-resistor interpretation of $\gamma$ requires this placement of the
forcing. Eq.~\eqref{eq:v} needs the recovery rate $\varepsilon(u,v)$ of
Eq.~\eqref{eq:eps}. An amplifier with a multiplier in its feedback path performs
the division. It forms $\varepsilon$ before $\varepsilon$ multiplies the recovery
bracket. On the attractor $u\ge-0.11$. The denominator therefore satisfies
$u+\mu_2\ge0.19$, and the division loop stays away from its singularity.

The summing amplifier has feedback resistor $R_f$, and each multiplier has the
scale factor listed in Table~\ref{tab:params}. The branch resistors reproduce the
coefficients of the bracketed term in Eq.~\eqref{eq:u}. At the values of
Table~\ref{tab:params} that term expands to
\begin{equation}
-8u^3+8.96u^2-0.96u-uv+0.4\cos\omega t .
\end{equation}
Circuit simulation verified the topology of Fig.~\ref{fig:schematic} before the
board was built.

\subsection{Amplifier Input-Stage Requirement}\label{sec:opamp}

The ideal model of Eqs.~\eqref{eq:u}--\eqref{eq:eps} says nothing about op-amp
input characteristics. The realization depends on them. The recovery-rate
divider is the resistive network that forms $\varepsilon$. It contains
$10$\,M$\Omega$ and $2$\,M$\Omega$ branches. The amplifier that senses this
network must draw negligible current. Its input impedance must also be well above
the branch resistances. The time constant of the divider is more than an order of
magnitude shorter than the forcing period. Divider bandwidth is therefore not the
limiting issue. Input loading is the relevant limitation, and component-level
simulation shows its size. A $\mu$A741-class bipolar input stage has a bias
current of order $10^{-7}$\,A and an input resistance of order a few megohms.
Such a stage injects current into the high-impedance divider and also shunts it.
In simulation, the $\mu$A741-class stage reproduces the large baseline attractor
reasonably well. At the rate-controlled setting it displaces the recovery node by
more than an order of magnitude. The failure is therefore most visible in the
small orbit.

The prototype uses TL084 JFET-input devices. Their input currents are at the
picoampere scale. Their input impedance is many orders of magnitude above the
divider resistances. With these devices the simulated circuit reproduced the
attractor of Eqs.~\eqref{eq:u}--\eqref{eq:eps} at $\gamma=0$. It also reproduced
the small orbit at the rate-controlled setting. The resulting design rule applies
to analog excitable models with a high-impedance, state-dependent rate network.
In such circuits that network sets the input-stage requirement of the amplifier.
The sub-hertz signal bandwidth does not.

\section{Measurements}\label{sec:meas}

\subsection{Prototype and Protocol}\label{sec:setup}

\begin{figure}[pos=htbp]
\centering
\includegraphics[width=0.6\linewidth]{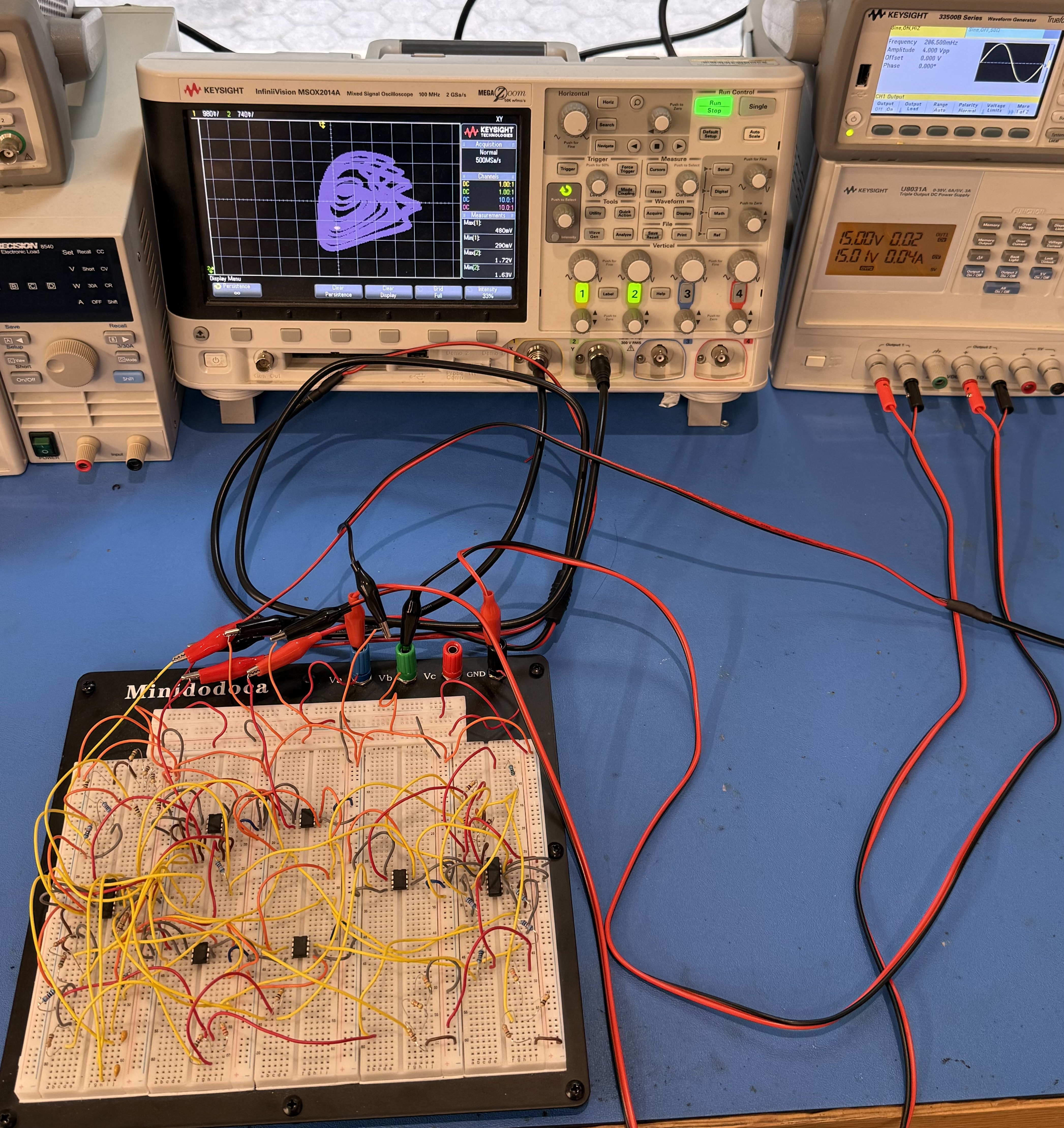}
\caption{Breadboard prototype and measurement setup. The analog computer occupies
a single breadboard panel. It contains two TL084 quad op-amp packages, five AD633
analog multipliers, and discrete $RC$ elements. The waveform generator at the top
right supplies the forcing at $286.500$\,mHz and $4.000$\,V$_{\mathrm{pp}}$. After
scaling, these values correspond to $\omega=1.8$\,rad\,s$^{-1}$ and $F=0.40$. The
triple-output supply provides $\pm15$\,V at $20$--$40$\,mA. The oscilloscope shows
the baseline attractor in $XY$ mode.}
\label{fig:bench}
\end{figure}

Fig.~\ref{fig:bench} shows the prototype and the measurement setup. The circuit
was built on a single breadboard panel with $\pm15$\,V rails and the component
values of Fig.~\ref{fig:schematic}. A Keysight 33500B generator supplied the
forcing, with its output-load setting at high impedance. A Keysight
InfiniiVision MSOX2014A oscilloscope recorded $V_U$ and $V_V$ in roll mode. The
records were exported as sampled waveform data. Table~\ref{tab:params} lists the
generator settings and the record length. The control resistors were measured,
not assumed. The six series resistors that form $R_u$ total $5.940$\,M$\Omega$,
and $R_0$ measures $1.01$\,M$\Omega$. By Eq.~\eqref{eq:Ru} the realized rate
parameter $\gamma=R_u/R_0-1$ is $4.881$, not the nominal $5$. The capacitor
$C_u$ measures $1\,\mu$F. All model predictions use the realized value of
$\gamma$.

Two independent runs were taken at each setting. Between runs the supply rails
were cycled, and the circuit settled for at least $120$\,s. That settling time
equals five tangent e-folding times $1/\lvert\lambda_1\rvert$ at the
rate-controlled setting. The two runs are thus separate experiments, not two
windows of one acquisition. Each measured quantity is the mean of the two runs.
Its uncertainty is the run-to-run standard deviation. Working from the sampled
record, instead of from instrument readouts, makes the estimator explicit.

\subsection{Excursion Estimators}\label{sec:estimator}

The excursion of a sampled record $x_n$ requires an estimator, and the two
settings need different ones. The peak-to-peak excursion is
\begin{equation}
\Delta_{\mathrm{pp}}=\max_n x_n-\min_n x_n .
\label{eq:dpp}
\end{equation}
The fitted excursion comes from a least-squares fit at the forcing frequency,
\begin{equation}
x_n=c_0+c_1\cos\omega t_n+c_2\sin\omega t_n+e_n ,
\label{eq:fit}
\end{equation}
where $t_n$ is the sample time, $c_0$, $c_1$, and $c_2$ are the fitted
coefficients, and $e_n$ is the residual. The fitted excursion is
\begin{equation}
\Delta_{\mathrm{f}}=2\sqrt{c_1^2+c_2^2}.
\label{eq:dfit}
\end{equation}

At the baseline setting the motion is aperiodic, and $\Delta_{\mathrm{pp}}$ is
appropriate. The baseline excursion spans more than a hundred quantization
levels, and additive noise is negligible against it. Less than a fifth of the
record power lies at the forcing frequency. At the rate-controlled setting the
orbit is periodic and locked. Almost all of the power of the $u$ record lies at
the forcing frequency. Most of the power of the $v$ record lies there too. The
excursion spans only tens of quantization levels. The noise extremes of the
record then bias $\Delta_{\mathrm{pp}}$ upward. The size of the bias depends on
the vertical scale in use. The two rate-controlled runs were recorded on purpose
at sensitivities that differ by a factor of four, so that the bias could be
measured. Between the two runs $\Delta_{\mathrm{pp}}$ differs by $11.8\%$, while
$\Delta_{\mathrm{f}}$ differs by $1.3\%$. The rate-controlled rows of
Table~\ref{tab:compare} therefore use $\Delta_{\mathrm{f}}$.

\subsection{Baseline and Rate-Controlled Behavior}\label{sec:results}

\begin{figure}[pos=!htbp]
\centering
\begin{subfigure}[b]{0.45\linewidth}
\includegraphics[width=\linewidth]{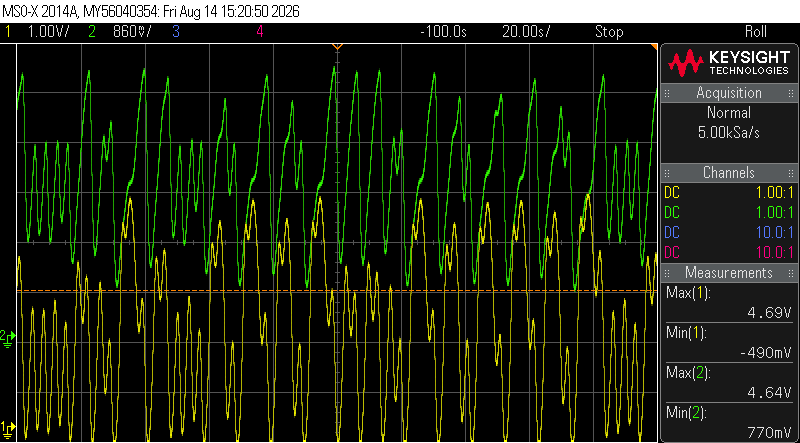}
\caption{}\label{fig:m-scope}
\end{subfigure}
\hfill
\begin{subfigure}[b]{0.47\linewidth}
\includegraphics[width=\linewidth]{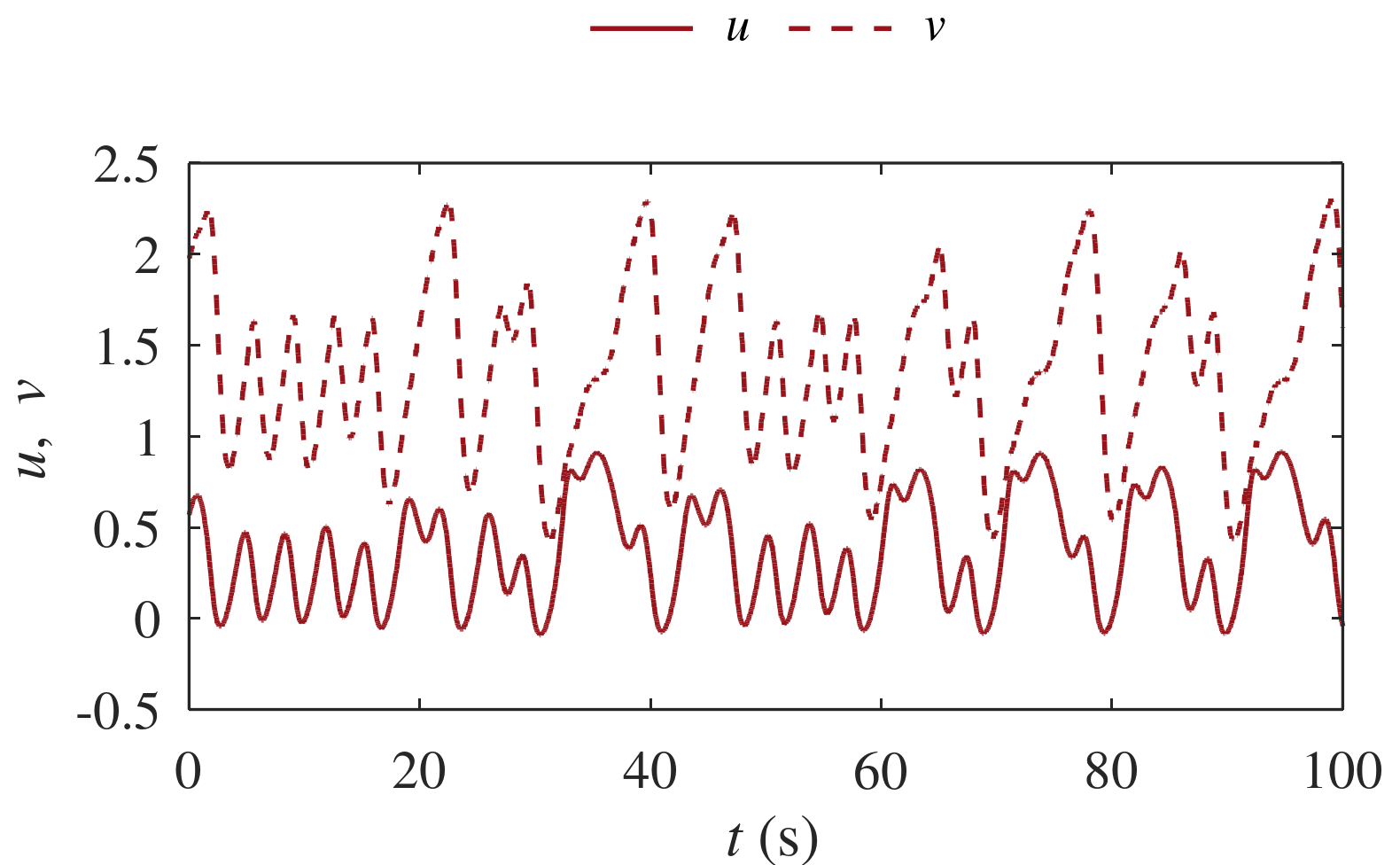}
\caption{}\label{fig:m-ts0}
\end{subfigure}\\[6pt]
\begin{subfigure}[b]{0.47\linewidth}
\includegraphics[width=\linewidth]{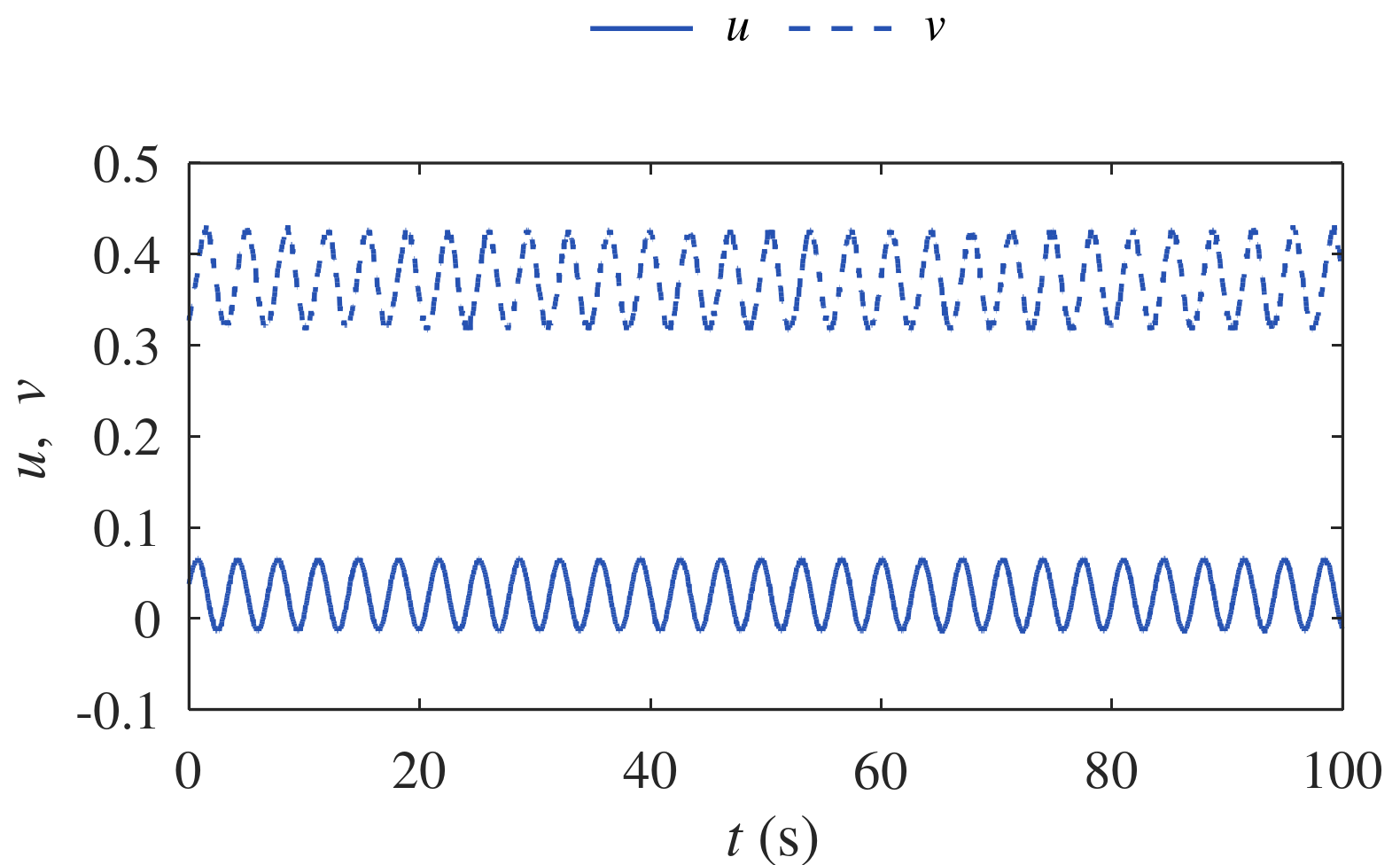}
\caption{}\label{fig:m-ts5}
\end{subfigure}
\hfill
\begin{subfigure}[b]{0.42\linewidth}
\includegraphics[width=\linewidth]{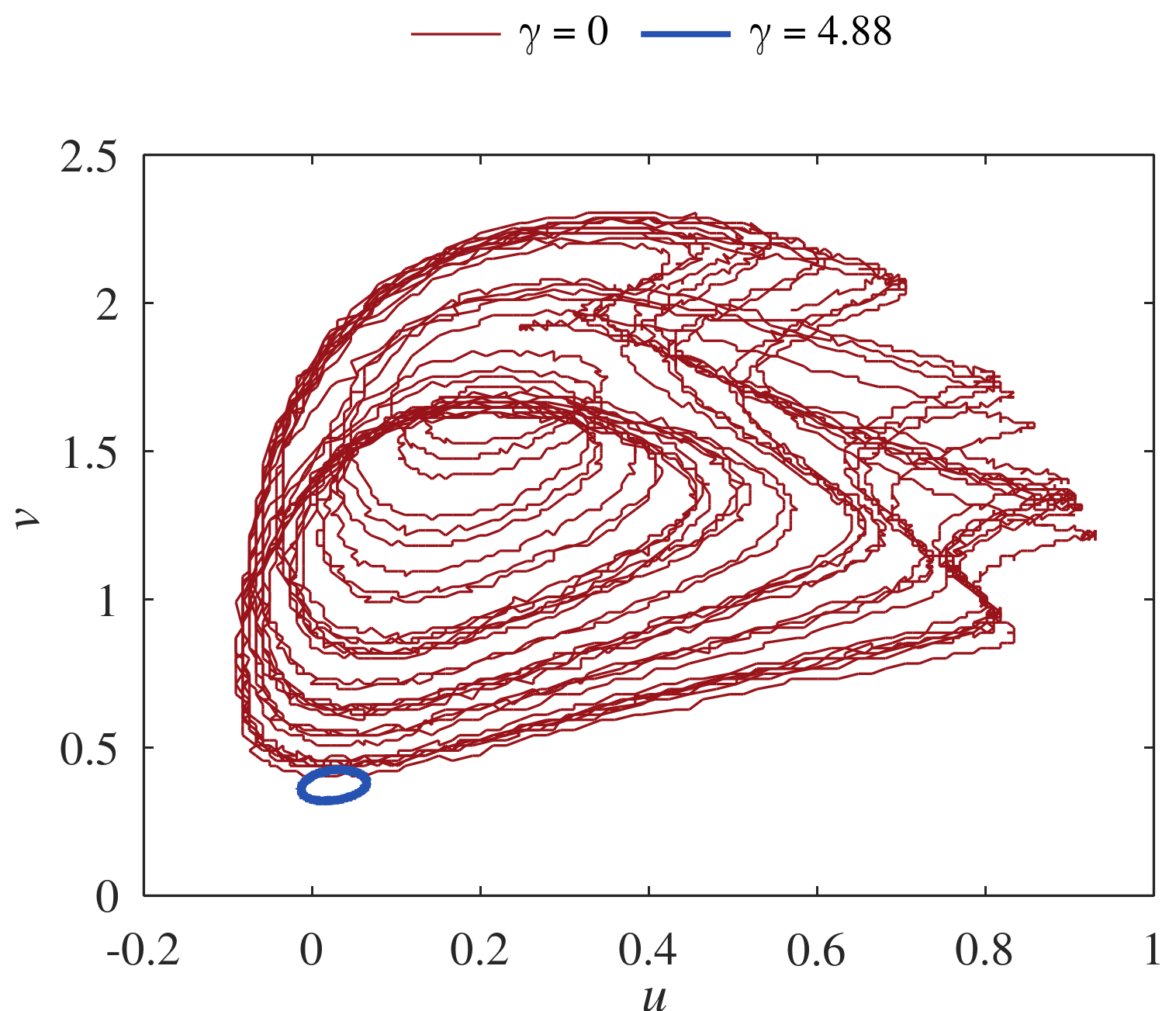}
\caption{}\label{fig:m-xy}
\end{subfigure}
\caption{Measured behavior of the prototype. Fig.~\ref{fig:m-scope} is the raw
$200$\,s baseline acquisition on the oscilloscope. Its yellow trace is channel 1,
$V_U$, and its green trace is channel 2, $V_V$. In Fig.~\ref{fig:m-ts0},
Fig.~\ref{fig:m-ts5}, and Fig.~\ref{fig:m-xy}, red denotes the baseline setting,
$R_u=1.01$\,M$\Omega$ and $\gamma=0$. Blue denotes the rate-controlled setting,
$R_u=5.940$\,M$\Omega$ and $\gamma=4.88$. Only this resistor changes.
Fig.~\ref{fig:m-ts0} and Fig.~\ref{fig:m-ts5} give sampled records at the two
settings over the first $100$\,s of $200$\,s. The excitation variable $u$ is
solid, and the recovery variable $v$ is dashed. The baseline records are
aperiodic over many forcing periods. The rate-controlled records are regular,
small, and locked to the forcing period. The two panels use independent vertical
scales, because the rate-controlled excursion is an order of magnitude smaller.
Fig.~\ref{fig:m-xy} gives the phase portraits of the two settings on common
axes over the full $200$\,s records. The baseline traces are from run 1, and the
rate-controlled traces are from run 2.}
\label{fig:measured}
\end{figure}

\begin{table}[pos=!htbp]
\caption{Measured and computed attractor excursions in model units at the
realized resistor ratios. Model units follow Eq.~\eqref{eq:scale}. Measured rows
give the mean $\pm$ standard deviation of two runs. Baseline rows use the
peak-to-peak excursion of Eq.~\eqref{eq:dpp}. Rate-controlled rows use the
fitted excursion of Eq.~\eqref{eq:dfit}.\label{tab:compare}}
\centering
\begin{tabular}{lccc}
\toprule
 & $\gamma$ & $u$ excursion & $v$ excursion \\
\midrule
Numerical integration & $0$ & $1.046$ & $1.913$ \\
Measured prototype    & $0$ & $1.017\pm0.006$ & $1.914\pm0.018$ \\
\midrule
Numerical integration & $4.88$ & $0.0753$ & $0.0131$ \\
Measured prototype    & $4.88$ & $0.0787\pm0.0007$ & $0.1030\pm0.0032$ \\
\bottomrule
\end{tabular}
\end{table}

Fig.~\ref{fig:m-scope} and Fig.~\ref{fig:m-ts0} show the baseline setting at
$R_u=1.01$\,M$\Omega$. Fig.~\ref{fig:m-scope} is the oscilloscope screen, and
Fig.~\ref{fig:m-ts0} is the sampled record in model units. Both channels are
aperiodic on the scale of many forcing periods. The red phase portrait in
Fig.~\ref{fig:m-xy} accumulates on a folded sheet with an inner recirculation
region and fanned outer bands. The orbit never closes on itself.
Table~\ref{tab:compare} compares the measured excursions with numerical
integration of Eqs.~\eqref{eq:u}--\eqref{eq:eps}. At the baseline setting they
agree to $2.8\%$ in $u$ and to $0.1\%$ in $v$.

Fig.~\ref{fig:m-ts5} shows the rate-controlled setting at
$R_u=5.940$\,M$\Omega$, with nothing else altered. The records become regular,
small in amplitude, and locked to the forcing period. The phase portrait
collapses to a single small closed orbit. Fig.~\ref{fig:m-xy} shows this blue
orbit to scale against the baseline attractor. The measured $u$ excursion agrees
with prediction to $4.5\%$. Between the two resistor values the $u$ excursion
contracts by a factor of $12.9$, against $13.9$ predicted. The residual few
percent corresponds to a forcing amplitude of about $4.2$\,V$_{\mathrm{pp}}$ at
the injection node, against the $4.000$\,V$_{\mathrm{pp}}$ set at the generator.
That difference is within the tolerance of the injection network.
Table~\ref{tab:compare} also shows one large deviation from the ideal model, and
it is reported as such. At the rate-controlled setting the measured $v$
excursion is nearly eight times the predicted value. The operating level of the
$v$ record is displaced from prediction by almost an order of magnitude.
Section~\ref{sec:slow} analyzes this deviation.

\section{Analysis and Discussion}\label{sec:analysis}

\subsection{Deviation in the Recovery Variable}\label{sec:slow}

The deviation is confined to the recovery path. It follows from the operating
level of that path, not from the rate-control mechanism. At the rate-controlled
setting the recovery node rests at $v=0.364$. Integration of
Eqs.~\eqref{eq:u}--\eqref{eq:eps} places it at $0.039$. The displacement is
$0.65$\,V at node V. The excitation node stays close to prediction. The
displaced operating level is enough to account for the excess $v$ excursion. The
recovery rate $\varepsilon(u,v)$ of Eq.~\eqref{eq:eps} depends on the state.
Evaluated at the measured operating point instead of the predicted one,
$\varepsilon$ rises by a factor of $6.6$. The measured $v$ excursion exceeds the
predicted one by a factor of $7.9$. The two factors are of the same size. The
excess $v$ excursion is therefore a consequence of the displaced level, not an
independent effect.

The displacement is visible only where the orbit is small. At the baseline
setting the same nodes agree closely, as Table~\ref{tab:compare} shows. A static
error in the recovery summing network is a fixed voltage. Such an error is a few
percent of a baseline excursion but several times a rate-controlled one. The
displacement also differs between the two settings. It is $0.15$ at the baseline
setting and $0.325$ at the rate-controlled setting. It also moves between bench
sessions. The captures reported here give $v=0.364$ at the rate-controlled
setting, and an earlier session on the same circuit gave $0.488$. A displacement
that varies with operating point and with session is not consistent with a fixed
component error. Thermal drift in the multiplier chain is consistent with it,
and the multipliers were observed to run warm. Leakage at the high-impedance
recovery-rate divider on the breadboard is also consistent with it. The present
data do not separate these two mechanisms. The recovery path is identified as the
location of the deviation, and its cause remains the open question. The
comparison with the model rests on the $u$ records. They agree to $2.8\%$ at the
baseline setting and to $4.5\%$ at the rate-controlled setting.

\subsection{Margin of the Rate-Controlled Setting}\label{sec:margin}

$R_u$ is the only control component. Its tolerance should not place the
prototype back near the chaotic boundary. At the measured $R_0$, a $\pm5\%$
tolerance about the nominal $6$\,M$\Omega$ spans $\gamma$ from roughly $4.6$ to
$5.2$. The exponent $\lambda_1$ was recomputed across that interval with the
settings of Table~\ref{tab:params}. It stays negative throughout. Its magnitude
stays within a fifth of its central value. The selected hardware point is
therefore not a knife-edge setting.

The margin at large $\gamma$ is not a monotone global suppression threshold. A
sweep of the same model at small $\gamma$ changes the sign of $\lambda_1$ more
than once. The sweep contains a narrow window of positive $\lambda_1$ below
$\gamma=1/4$. The hardware point is placed well above this transition region. At
the rate-controlled setting, independent changes of $\pm15\%$ in $F$ and in
$\omega$ leave $\lambda_1$ negative in every case. The design thus has useful
margin against generator setting errors.

\subsection{Scope and Limitations}\label{sec:disc}

The values of $\lambda_1$ in Table~\ref{tab:params} certify the two operating
points of the model. The oscilloscope records are not used by themselves as a
proof of chaos. A finite noisy record can make quasiperiodic motion look
irregular. The hardware establishes that the resistor ratios of the circuit
implement Eqs.~\eqref{eq:u}--\eqref{eq:eps}. The baseline excursions agree
closely with the computed values in Table~\ref{tab:compare}. The measured phase
portrait collapses to a single closed orbit when only $R_u$ changes. The
exponent $\lambda_1$ of the model is positive at $R_u=1$\,M$\Omega$ and negative
at $R_u=6$\,M$\Omega$. This sign change makes the chaotic-to-regular terminology
quantitative without over-interpreting the finite hardware record.

Breadboard leakage at the recovery-rate divider is a further limitation. The
comparison of TL084 and $\mu$A741-class stages bears on this limitation. The
comparison identifies a failure mechanism that the ideal equations do not show.
Leakage at this divider is one of the two candidate causes of the operating-point
displacement identified in Section~\ref{sec:slow}. Thermal drift in the
multiplier chain is the other. A guarded printed-circuit layout with shorter
high-impedance traces is the appropriate next step for precision amplitude
comparison. Measured component values would replace nominal ones in that step.
The guarded layout would remove the leakage path. It would then test the leakage
explanation without changing the control concept.

Changing $R_u$ alone moves the tested circuit from the chaotic setting to a
regular one. The change needs no delay element, no digital loop, and no added
state. It is \emph{rate control}, not state feedback. Unlike delayed feedback,
rate control does not preserve an embedded unstable periodic orbit of the
baseline system. It applies only where a common gain multiplies the complete
right-hand side of the excitation equation. The transition must therefore be
verified for each model and operating point, not assumed.

\section{Conclusion}\label{sec:concl}

A single passive resistor takes the analog forced Aliev--Panfilov oscillator out
of its chaotic regime. The contributions of the paper, with their measured
outcomes, are the following.
\begin{itemize}
\item Chaos control by one resistor. Changing the input resistor of the
excitation integrator from $1$\,M$\Omega$ to $6$\,M$\Omega$ moves the circuit
from the chaotic to the regular operating point of the model. The largest
Lyapunov exponent of the model changes sign between the two settings. No state,
delay element, or digital controller is added. In the model, the setting remains
regular under a $\pm5\%$ resistor tolerance and $\pm15\%$ forcing errors.
\item Measured validation. The $u$ records agree with numerical integration to
$2.8\%$ at the baseline setting and to $4.5\%$ at the rate-controlled setting,
realized at $\gamma=4.88$. Between the two settings the $u$ excursion contracts
by a factor of $12.9$, against $13.9$ predicted. Two independent runs support
each setting, and the sampled records are provided.
\item An amplifier design rule. The state-dependent recovery-rate divider
requires an amplifier with very high input impedance. TL084 JFET-input devices
avoid the displacement seen with a tested $\mu$A741-class bipolar stage.
\item Localization of the remaining discrepancy. At the rate-controlled setting
the $v$ excursion remains eight times its ideal prediction. The excess follows
from a displaced operating point on the state-dependent recovery rate, not from
the excitation channel.
\end{itemize}

The displacement in the recovery path moves between bench sessions. Thermal
drift in the multiplier chain and leakage at the recovery-rate divider could
each produce it. Separating the two is the open question. The Aliev--Panfilov
equations originate as a phenomenological description of cardiac
excitation~\cite{aliev}. A compact analog realization of these dynamics is
useful as a low-power hardware testbed. Extending the single-resistor mechanism
to arrays of coupled cells is the step required before spatially extended
excitable media can be addressed. The rate parameter $\gamma$ has no direct
physiological counterpart. This work implies no therapeutic application.

\section*{Acknowledgments}
The circuit was built and all hardware measurements were made in
the NYU Tandon MakerSpace, Brooklyn, NY 11201. AI tools were used for editorial
assistance, mathematical cross-checking, and language refinement. All theoretical developments, derivations, conclusions 
and figures are the authors' work.

\printcredits

\section*{Declaration of competing interest}

The authors declare no competing interests.

\section*{Funding}

No funding received.

\section*{Data availability}

The sampled waveform records that support the findings of this study are
provided as supplementary data files. There are two records at each resistor
setting, each holding $6400$ samples of $V_U$ and $V_V$ over $200$\,s.

\end{document}